\documentclass[twocolumn,amsmath,amssymb,aps,pre,reprint,longbibliography]{revtex4-1}
\usepackage{graphicx}
\begin{document}

\title{
Phase Separation, Edge Currents, and Hall Effect for Active Matter with Magnus Dynamics 
} 

\author{
B. Adorj{\' a}ni$^{1}$, A. Lib{\' a}l$^{1}$, C. Reichhardt$^{3}$, and C. J. O. Reichhardt$^{3}$
}
\affiliation{
$^{1}$ Mathematics and Computer Science Department, Babe{\c s}-Bolyai University, Cluj 400084, Romania\\
$^{2}$ Theoretical Division and Center for Nonlinear Studies,
Los Alamos National Laboratory, Los Alamos, New Mexico 87545, USA}
\date{\today}

\date{\today}
\begin{abstract}
  We examine run and tumble disks in two-dimensional systems where the particles also have a Magnus component to their dynamics. For increased activity, we find that the system forms a motility-induced phase-separated (MIPS) state with chiral edge flow around the clusters, where the direction of the current is correlated with the sign of the Magnus term. The stability of the MIPS state is non-monotonic as a function of increasing Magnus term amplitude, with the MIPS region first extending down to lower activities followed by a break up of MIPS at large Magnus amplitudes into a gel-like state. We examine the dynamics in the presence of quenched disorder and a uniform drive, and find that the bulk flow exhibits a drive-dependent Hall angle. This is a result of the side jump effect produced by scattering from the pinning sites, and is similar to the behavior found for skyrmions in chiral magnets with quenched disorder. 
\end{abstract}
\maketitle

\section{Introduction}

There is growing interest in active matter systems or systems where some form of self-motility occurs, which are relevant in soft matter, robotic,
and biological settings
\cite{Marchetti13}.
One of the most well-known effects observed in
collections of interacting active particles is
motility-induced phase separation (MIPS)
\cite{Fily12,Redner13,Palacci13,Buttinoni13,Cates15,Reichhardt14a}.
Here, even repulsive disks that would form a
uniform fluid in thermal equilibrium
instead assemble into an active solid
that coexists with a low density active liquid.
Recent work on active matter systems has increasingly focused on
how active matter behaves when
coupled to a complex environment, such as confinement,
mixtures of active and inactive particles,
or a disordered substrate that can
enhance or disrupt the MIPS clustering depending on the
substrate interaction strength.
\cite{Chepizhko13,Morin17,Sandor17b,Yllanes17,Sandor17a,Chardac21}.
Different types of trapping effects can also arise
\cite{Bhattacharjee19a,Shi20}. 
When active matter interacting with a disordered landscape is
subjected to bulk driving,
depinning phenomena can occur similar to
that found for driven passive particles interacting with random disorder.
In some cases, increasing the activity can reduce
the net mobility through the system
due to the enhancement of self-trapping by the activity \cite{Reichhardt14}.

Most active matter systems obey overdamped dynamics; 
however, if additional chiral dynamics are present,
such as due to the spinning of the particles,
a non-dissipative Magnus term arises.
The Magnus term
generates a velocity component perpendicular to the
net forces experienced by the particle,
and Magnus effects appear for fluid vortices
\cite{Aref88,Reichhardt20b},
magnetic vortices \cite{Novosad05},
charges in a magnetic field \cite{Melzer21,Shinde22},
fracton phases \cite{Doshi21},
and skyrmion textures in chiral magnets
\cite{Muhlbauer09,Nagaosa13,EverschorSitte14,Reichhardt22a}.
Particles with a Magnus term that are subjected to uniform driving
move at a Hall angle $\theta_H$ with respect to the driving direction,
and $\theta_H$ increases as the Magnus term becomes larger.
Magnus-induced Hall angles
similar to the Hall effect found for charges moving in
a magnetic field \cite{Hall1879}
have been observed in simulations
\cite{Reichhardt22a,Reichhardt15,Lou22}
and experiments \cite{Jiang17,Litzius17}
for skyrmions in chiral magnets.
The size of the Hall angle at low drives is reduced in the presence
of quenched disorder 
due to a side jump effect in which particles are scattered
sideways as they cross a pinning site, but for high drives this
scattering is reduced and
the Hall angle
approaches its intrinsic substrate-free value
\cite{Reichhardt15,Jiang17,Litzius17,Reichhardt22a}.

Magnus effects can arise in active matter systems when the
particles have a chiral shape \cite{Kummel13} or are
undergoing spinning or circular motion
\cite{Lemelle10,Nourhani15,Lowen16,Tan17,Liebchen22}.
Chiral active colloids have been shown to produce
edge currents and odd viscosity effects
\cite{vanZuiden16,Banerjee17,Dasbiswas18,Soni19,Reichhardt19,Yang20,Han21,Fruchart23,Petroff23}, while
other studies have demonstrated that chirality can increase the frequency
of scattering events and
thus increase the diffusive motion of the particles \cite{Kalz22}.
Recently, Cao {\it et al.} produced an experimental realization of
spinning colloids in viscoelastic fluids that generated
very strong Magnus forces and a pronounced Hall angle
\cite{Cao23}.
For individual particles
driven through chiral fluids,
a drive dependent Hall angle has been observed
\cite{Reichhardt19a}, and it was also shown that in some cases,
a velocity boost effect occurs in which the
velocity increases due to the conversion of some of
the rotational velocity into translational velocity
by the Magnus term
\cite{Reichhardt22b,Poggioli23}.
Chiral active matter has also been shown to form
large rotating crystals \cite{Nguyen14,Aubret18,Tan22} and to
exhibit
wave propagation due to odd viscosity effects \cite{Tan22}.
Biological systems often involve chiral active
matter, such as in 
the chiral motion of bacteria \cite{DiLuzio05}                  
or the formation of bacterial vortex lattices \cite{Wioland16,Reinken20}.

Here we explore how the behavior of active matter changes when a
Magnus term is added to the equations of motion.
We find that in the MIPS regime,
a system with finite Magnus forces forms
a rigidly rotating crystalline cluster surrounded by a fast-moving circulating
edge current of particles.
As the Magnus force increases, the edge current region develops
additional shear bands and contains
multiple rows of particles moving at different velocities.
We also find that the Magnus term stabilizes the MIPS
clusters down to lower running lengths compared to Magnus-free systems.
This occurs because forces directed radially away from the cluster, which
ordinarily would tend to break the cluster apart, are rotated
by the Magnus term into the
direction perpendicular to the cluster edge, establishing an edge current
and keeping the particles localized at the cluster.
For very strong Magnus forces, the clusters break up into chiral labyrinthine
states.
We examine the dynamics when a random pinning landscape and a drift force
are introduced.
In the absence of quenched disorder, the clusters
move at a constant Hall angle with respect to the
direction of the drift force; however, when quenched disorder is present,
the Hall angle develops a drive dependence, rising from a low value
at low drives before eventually approaching the disorder-free value at
high drives.
This drive dependence of the Hall angle occurs due to the
Magnus-induced asymmetric scattering of active
particles by the pinning sites, which 
is strongest at the lowest drives
and decreases for higher drives
when the particles spend less time interacting with
any given pinning site.
A similar drive-dependent Hall angle for motion
over quenched disorder has been observed for
magnetic skyrmions, which also have a strong Magnus term
in their dynamics.
We find that the MIPS cluster breaks apart
when the disorder is strong,
but that the cluster can partially reform at
higher drives when the particles are moving sufficiently rapidly that
the pinning effectiveness is reduced.
We also show that a drive dependent Hall angle can occur for
particles with a Magnus term but no activity. Our results should be relevant to
chiral active matter systems and systems with odd viscosity.

\section{Simulation}

We model a two-dimensional system
of size $L \times L$ with $L=160$ that has
periodic boundary conditions in both the $x$ and $y$ directions. 
The sample contains $N = 4000$ circular disks of radius 
$R_{d} = 1.0$, giving a particle density of 
$\phi=N\pi R_d^2/L^2 = 0.49$.
Each particle moves under the influence of
a motor force ${\bf f}^{\rm mot}_i=F_m{\hat {\bf m}}_i$ 
of magnitude $F_m = 1.0$ applied
in a randomly chosen direction
${\hat {\bf m}}_i$ during a
time interval of $\tau_{i}$
simulation time steps, drawn randomly
from the allowed range $\tau_i \in [\tau, 2\tau]$.
After $\tau_i$ time steps have elapsed,
particle $i$ undergoes an instantaneous
tumbling event in which new random
values of $\tau_i$ and ${\hat {\bf m}}_i$ are selected.
Particles in contact with each other experience a repulsive
harmonic potential
${\bf f}^{\rm dd} = f_{\rm dd}{\bf \hat{r}}_{ij}$ with $f_{\rm dd}=\sum_j^N k (2R_{d}-r_{ij})\Theta(2R_d-r_{ij})$,
where $r_{ij}=|{\bf r}_j-{\bf r}_i|$ is the distance between particles
located
at ${\bf r}_i$ and ${\bf r}_j$,
${\hat {\bf r}}_{ij}=({\bf r}_j-{\bf r}_i)/r_{ij}$,
$\Theta$ is the Heaviside step function, and $k = 20$
is the elastic constant.

The overdamped equation of motion for particle $i$ is

\begin{equation}
\alpha_d {\bf v}_{i} + \omega{\bf {\hat z}} \times {\bf v}_i =
{\bf f}^{\rm dd}_{i} + {\bf f}^{\rm mot}_{i} + {\bf f}^{\rm pin}_{i}
+ {\bf f}^{D} \ .
\end{equation}

where ${\bf v}_{i} = {d {\bf r}_{i}}/{dt}$ is the particle velocity.
The first term on the left hand side
is the damping term of magnitude $\alpha_d$,
while the second term is the Magnus force
of magnitude $\omega$, which contributes
a velocity component perpendicular to the net force on the particle.
Interactions with $N_{\rm pin}$ randomly placed nonoverlapping
parabolic pinning sites of radius $R_{\rm pin}=0.5$ and maximum
strength $F_p$
are given by
${\bf f}^{\rm pin}_i = \sum_k^{N_{\rm pin}} F_{\rm p} (r_{ik}^{(p)}/R_{\rm pin}) \Theta(r_{ik}^{(p)} - R_{\rm pin})\hat{\bf r}_{ik}^{(p)}$,
where $r_{ik}^{(p)} = |{\bf r}_i - {\bf r}_k^{(p)}|$ is the distance between
particle $i$ and the pinning site at ${\bf r}_k^{(p)}$, and
$\hat{\bf r}_{ik}^{(p)}  = ({\bf r}_{k}^{(p)} - {\bf r}_{i})/r_{ik}^{(p)}$.
The final term, ${\bf f}^{D}=F_D{\bf \hat{x}}$,
represents a uniform drift force applied to all of the particles.

\begin{figure}
\includegraphics[width=\columnwidth]{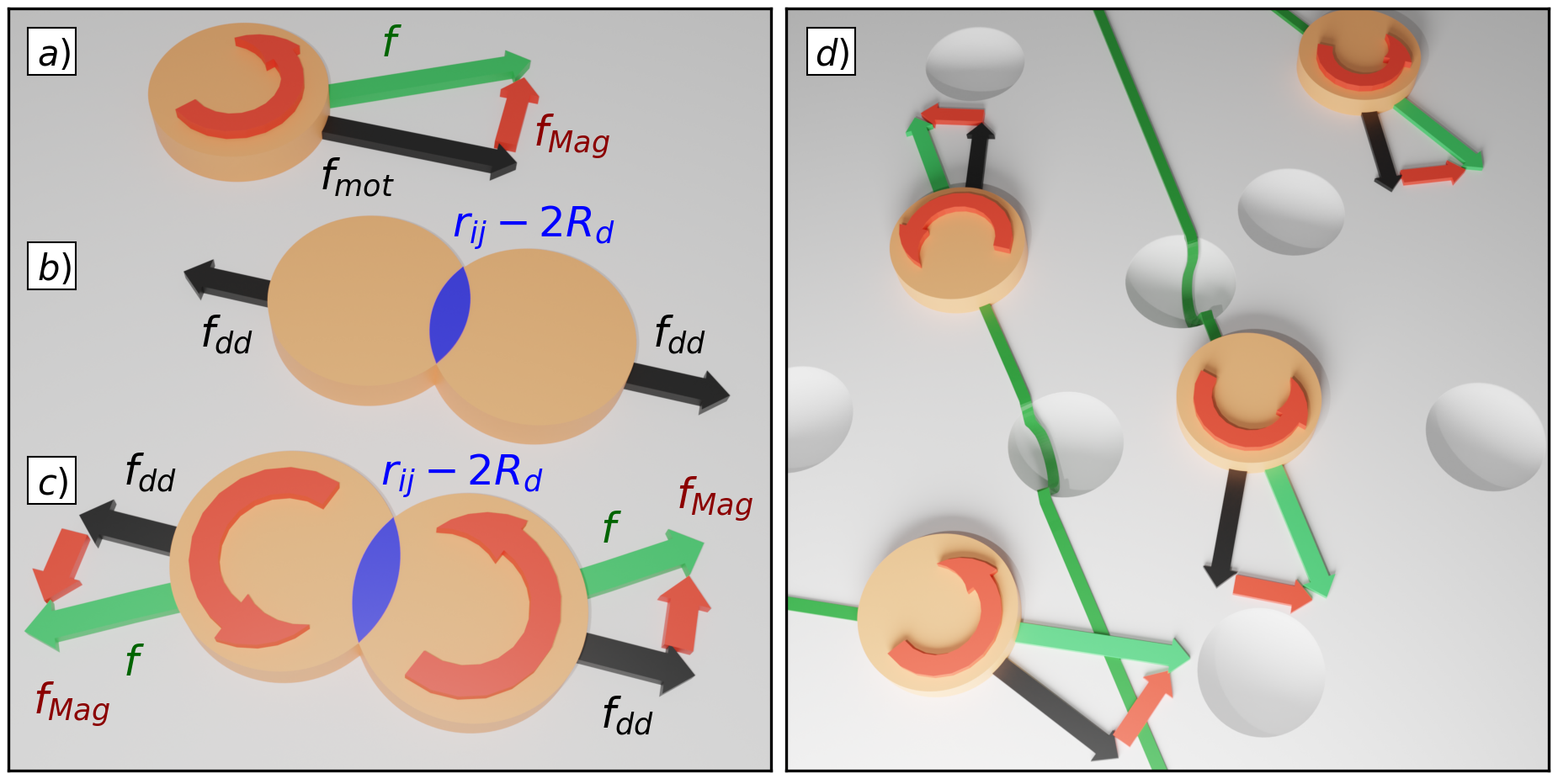}
\caption{
(a) Schematic of the forces acting on the run-and-tumble
Magnus particles (tan disks). The intrinsic propulsion arises from the
motor force $f^{\rm mot}$ (black arrow).
The Magnus force $f^{\rm Mag}$ (straight red arrow) acts perpendicularly to
the net forces on the particle, which in the absence of pinning or
particle-particle interactions is only the motor force,
giving a final force of $f$ (green arrow). The Magnus force effectively
rotates the net forces counterclockwise, as indicated by the circular red
arrow.
(b) In the absence of the Magnus force, two particles with motor forces
directed toward each other (not shown) experience a contact repulsion
$f^{\rm dd}$ (black arrows) proportional to the net overlap
$r_{ij}-2R_d$, where $R_d$ is the particle radius.
(c) When a Magnus force $f^{\rm Mag}$ (straight red arrow)
is added to the particles from panel (b),
it produces final forces $f$ (green arrows) that are rotated counterclockwise
(circular red arrows).
(d)  Schematic of an assembly of run-and-tumble Magnus particles (tan disks)
interacting with randomly placed non-overlapping pinning sites
(divots). Black arrows indicate the instantaneous motor force of each
particle, green arrows show the net forces that have been rotated by
the Magnus forces (red arrows), and thin green lines illustrate the
trajectories of the particles.}
\label{fig:1}
\end{figure}

In Fig.~\ref{fig:1}(a) we show a schematic of the effect of the Magnus
term on a freely moving active particle. The Magnus term acts perpendicular
to the net applied force, which is only the motor force in this example,
and effectively rotates the force counterclockwise, producing a finite Hall
angle between the direction of force and the direction of motion.
The elastic interaction between two particles with oppositely directed
motor forces is illustrated in
Fig.~\ref{fig:1}(b), where the net force pushes the particles directly away
from each other. When a Magnus term is added to this interaction,
Fig.~\ref{fig:1}(c) shows that the repulsive forces are effectively
rotated counterclockwise, causing the particles to be pushed away from
each other at an angle to the contact interaction force.
In Fig.~\ref{fig:1}(d) we plot a schematic of the motion of a collection
of active Magnus particles driven over randomly placed pinning sites.
Upon traversing a pinning site, the particles experience a side slip
generated by the Magnus term that reduces the size of the Hall angle. The
magnitude of this side slip diminishes as the external drift force $F_D$
increases, causing the Hall angle to approach its intrinsic value
at high drives.

To characterize the behavior, we measure the average fraction $C_L$ of
particles in the largest cluster, determined using the algorithm
described in Ref.~\cite{Luding99}.
When pinning and a drift force are both present, we also
measure the average transport parallel,
$\langle V_{x}\rangle = N^{-1}\sum^{N}_{i}{\bf v}_i\cdot {\bf \hat{x}}$,
and perpendicular,
$\langle V_{y}\rangle = N^{-1}\sum^{N}_{i}{\bf v}_i\cdot {\bf \hat{y}}$,
to the applied drive.
The measured Hall angle is given by
$\theta_H  = \arctan(\langle V_{y}\rangle/\langle V_{x}\rangle)$. 
In the absence of particle-particle interactions or pinning,
the intrinsic Hall angle is 
$\theta_H^{\rm int} = \arctan(\omega/\alpha_{d})$.
We use a simulation time step 
of size $dt = 0.001$ and
allow the system to evolve
to a steady state during $2.5\times 10^6$ simulation time steps
before collecting data during the next $2.5\times 10^6$ simulation time steps.

\section{Results}

\subsection{No Quenched Disorder or Drift}

\begin{figure}
\includegraphics[width=\columnwidth]{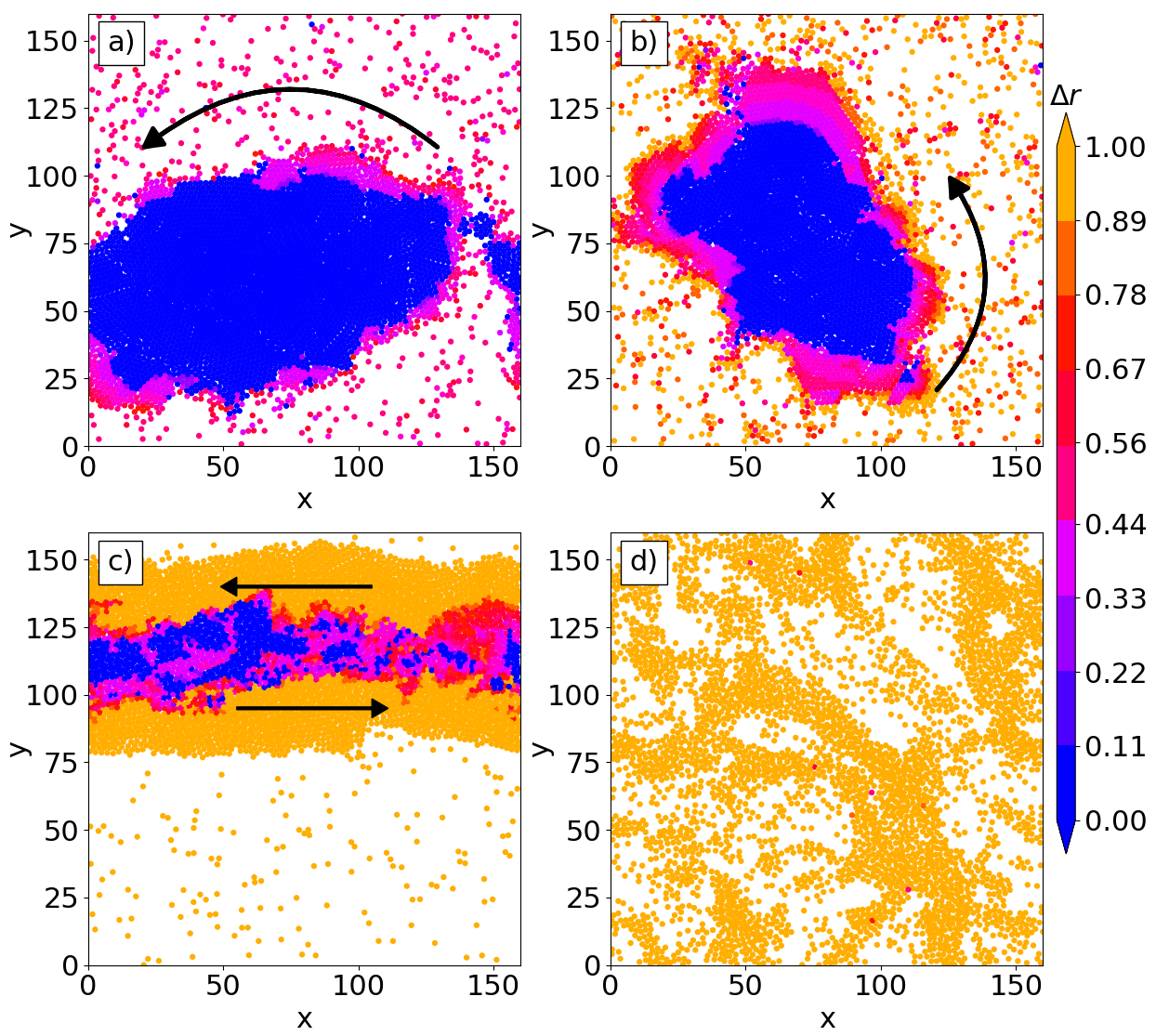}
\caption{Images of particle configurations
in samples with active Magnus particles for
zero pinning and zero drift force, $N_{\rm pin}=0$ and $F_D=0$,
at a particle density of $\phi=0.49$ and a run time 
$\tau = 1.4\times 10^5$ for different Magnus term magnitudes
of $\omega =$
(a) 1, (b) 2, (c) 6, and (d) 10.
Colors indicate the net displacement $\Delta r$ of the particle
during a period of 500 simulation time steps.
Arrows indicate the rotation direction of the cluster in panels (a) and (b),
and the direction of shear banding flow in panel (c).
} \label{fig:2}
\end{figure}

We first consider the behavior in the absence of pinning or a drift
force in a sample with $\phi = 0.49$ and $\tau=1.4\times 10^5$.
For these parameters, a MIPS state appears when
the Magnus term is absent, $\omega=0$.
Images of the instantaneous particle positions
appear in Fig.~\ref{fig:2}(a-d) for Magnus term magnitudes of
$\omega=1$, 2, 6, and 10, respectively. The particles are colored according
to their net displacement $\Delta r=|{\bf r}_i(t)-{\bf r}_i(t-\Delta t)|$
during a time period of $\Delta t=500$ simulation time steps.
For the small but finite Magnus term of $\omega=1$ in Fig.~\ref{fig:2}(a),
we find a MIPS state where the dense cluster has a
net rotation, indicated by an arrow,
that is in the same direction as the Magnus force chirality.
The rotation is not uniform throughout the cluster; instead, there is
a dense nonrotating core surrounded by a rotating halo of particles
at the edge of the cluster.
The halo particles move from three to eight times faster than
particles in the nonrotating core, and as a result, a shear band forms
near the cluster edge.
For $\omega = 2$ in Fig.~\ref{fig:2}(b), the rotation speed of the
cluster edge has increased and a velocity gradient has developed
inside the cluster edge, leading to the appearance of
multiple shear band regions.
Here, the dense core region of the cluster rotates slowly as a rigid object.
The cluster becomes more elongated as the Magnus term increases,
and in
Fig.~\ref{fig:2}(c) at $\omega = 6$ the cluster spans the width of the system
and develops wide edge regions that are rapidly moving in opposite directions
on the top and bottom of the cluster.
Here, even though a large dense cluster still forms, the strong shearing
forces destroy the crystalline ordering in the cluster core.
When the Magnus term becomes sufficiently large, it interferes with the
ability of the particles to nucleate stable clusters due to the force
rotation effect illustrated in Fig.~\ref{fig:1}(c), and the system
breaks apart into the clustered fluid state illustrated
in Fig.~\ref{fig:2}(d) for $\omega=10$.

\begin{figure}
\includegraphics[width=\columnwidth]{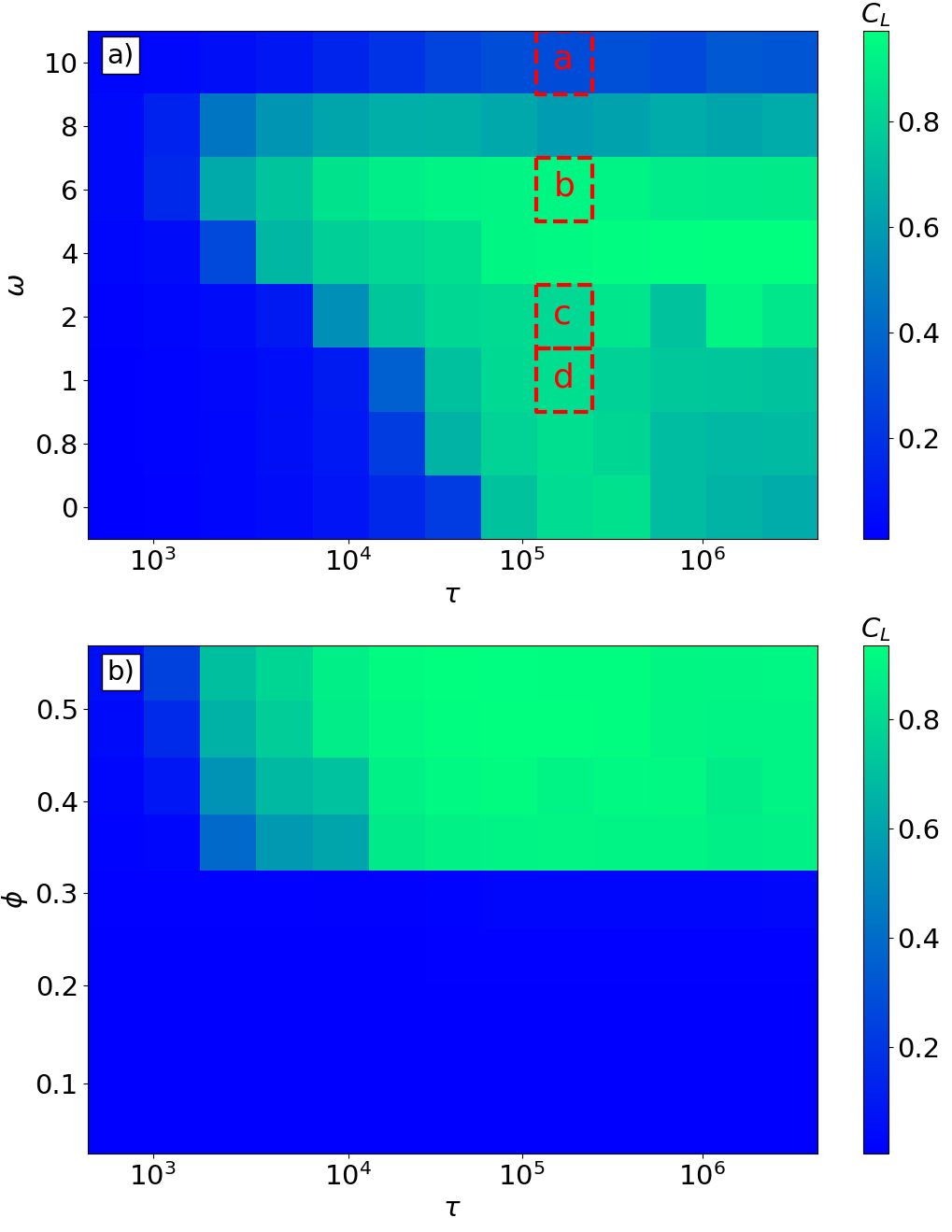}
\caption{Heat maps of the fraction $C_L$ of particles in the largest
cluster for the system from Fig.~\ref{fig:2} with $N_{\rm pin}=0$ and
$F_D=0$.  
(a) $C_L$ as a function of $\omega$ vs $\tau$ at $\phi = 0.49$.
Squares indicate the values of $\tau$ and $\omega$ at which the images
in Fig.~\ref{fig:2} were obtained.
As $\omega$ increases up to $\omega=6$, the window of $\tau$ for
which strong clustering occurs becomes wider.
Edge currents only appear when $\omega > 0$.
(b) $C_{L}$ as a function of $\phi$ vs $\tau$ at $\omega = 6$
showing that a MIPS state cannot form when $\phi\leq 0.31$.
}
\label{fig:3}
\end{figure}

In Fig.~\ref{fig:3}(a), we plot the fraction $C_L$
of particles in the largest cluster $C_L$
as a function of $\omega$ versus $\tau$ for the
system from Fig.~\ref{fig:2} with $N_{\rm pin}=0$ and $F_D=0$.
Dashed squares highlight the values of $\omega$ and $\tau$ at which
the images in Fig.~\ref{fig:2} were obtained.
When $\omega = 0$, the system forms
an ordinary nonrotating MIPS state when $\tau$ is sufficiently large,
as observed in previous studies
\cite{Fily12,Redner13,Cates15,Bechinger16}.
As $\omega$ increases to $\omega=6$,
the region where MIPS clusters appears expands 
by more than a factor of 10 to lower values of $\tau$,
indicating that the Magnus term is helping to stabilize the MIPS state.
At $\omega = 0$, particles near the cluster edge remain attached to
the cluster only as a result of their motor force, which must be
aligned in the direction of the cluster in order to prevent a particle
from escaping the cluster.
When $\omega$ is finite, a motor force directed away from the cluster
gets rotated and becomes more perpendicular to the cluster edge,
causing particles that would have escaped from the cluster to
start rotating around the cluster instead,
generating an edge current and helping
to stabilize the cluster down to lower values of $\tau$.
In previous work on driven skyrmions, it was shown that the Magnus force
tended to stabilize skyrmion clusters even though the skyrmion-skyrmion
interactions are strictly pairwise repulsive,
and that this enhanced clustering originated when the skyrmions tended
to spiral around each other instead of moving away from each other
\cite{Reichhardt22a}.
In Fig.~\ref{fig:3}(b), we plot
a heat map of $C_{L}$ as a function of $\phi$ versus
$\tau$ for the same system from
Fig.~\ref{fig:3}(a) at $\omega = 6$.
Similar to ordinary MIPS systems with $\omega=0$, both 
the activity and the density must be high enough or else the MIPS
state does not appear; however, the onset of MIPS
occurs at $\tau =2\times 10^3$ when $\phi = 0.49$, which is significantly
lower than the onset value of $\tau=7 \times 10^5$ found for
$\omega=0$ in Fig.~\ref{fig:3}(a).
We find in Fig.~\ref{fig:3}(b) that a MIPS
cluster only appears if $\phi > 0.31$.

Our results can be compared to several experiments.
In Ref.~\cite{vanZuiden16},
spinning colloids that form cluster states develop edge currents
similar to what we observe.
The giant rotating crystal that we find resembles states
seen in
spinning starfish embryos, which form rigidly rotating clusters
\cite{Tan22}.
Unlike the embryos, which have an
intermediate-range attractive component in the
particle-particle interactions due to hydrodynamic effects, for our system
the rotation is not rigid but instead produces a
shear band that separates a rapidly rotating edge region
from a more slowly rotating or nonrotating bulk rigid crystal region,
as shown in Fig.~\ref{fig:2}(a,b).
The particle-particle interactions in our system are strictly
repulsive and the cluster can only be stabilized by the motor force and
by the Magnus rotation of the motor force.

\subsection{Quenched Disorder Plus Drift}

\begin{figure}
\includegraphics[width=\columnwidth]{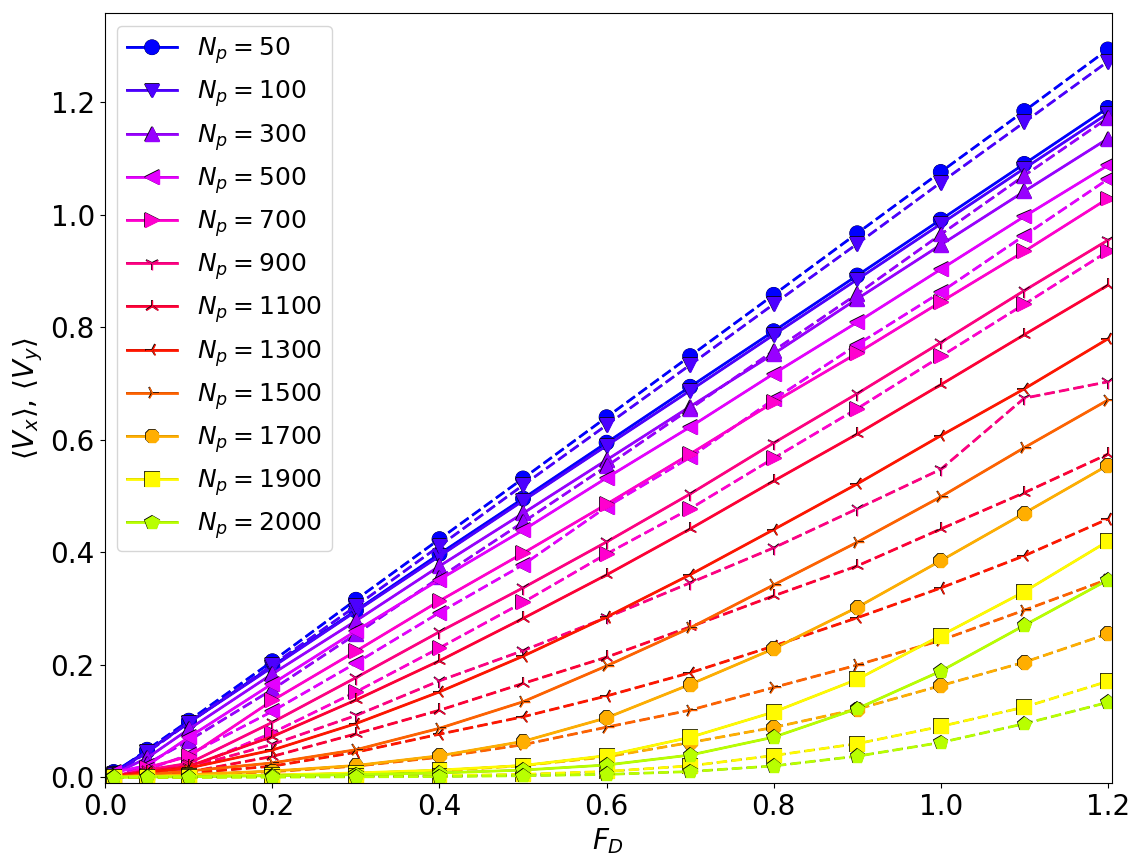}
\caption{
$\langle V_{x} \rangle$ (solid lines) and $\langle V_{y} \rangle$
(dashed lines) vs $F_{D}$ in a sample with $\omega=1.1$, $\phi=0.49$,
and $\tau=1.4\times 10^5$ 
at $F_p=7$ for varied number of pinning sites $N_p=50$, 100, 300, 500,
700, 900, 1100, 1300, 1500, 1700, 1900, and 2000, from top to bottom.
}
\label{fig:4}
\end{figure}

\begin{figure}
\includegraphics[width=\columnwidth]{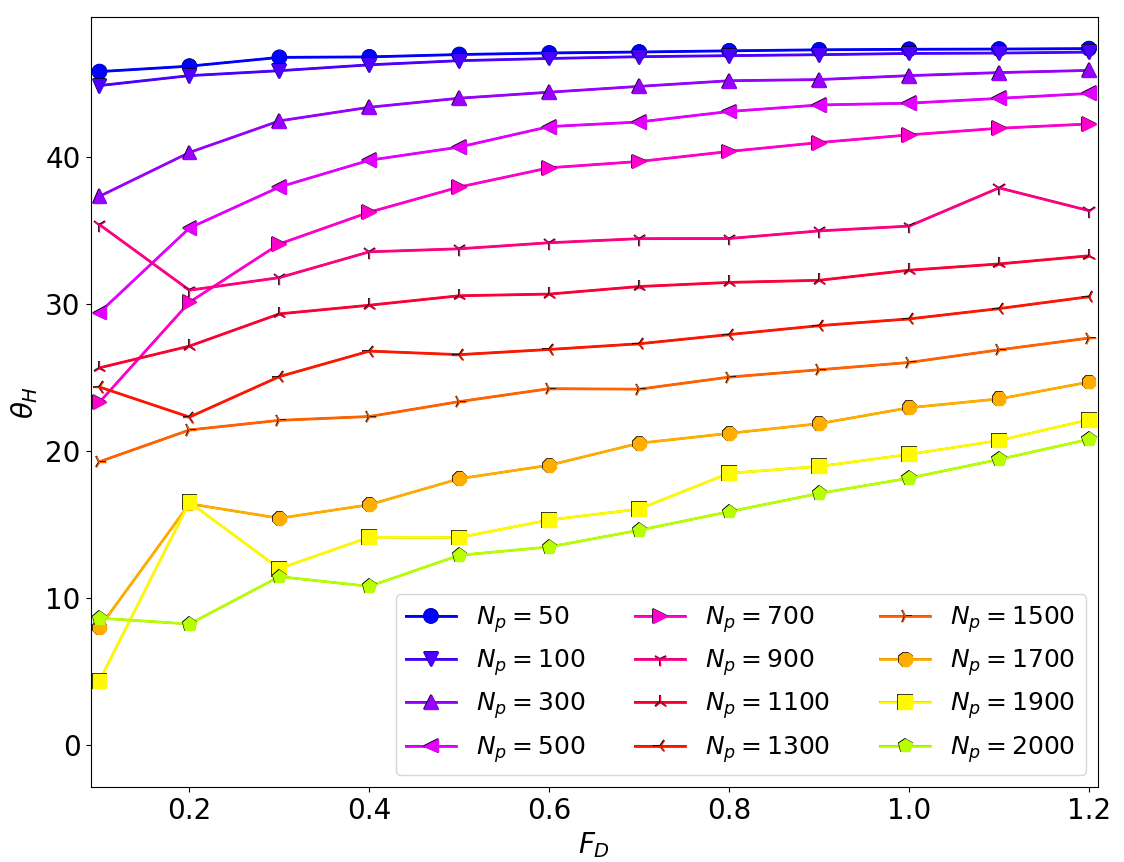}
\caption{
The measured Hall angle $\theta_H  = \arctan(\langle V_y \rangle/\langle V_x \rangle)$ vs $F_{D}$ for the
system from Fig.~\ref{fig:4} with $\omega=1.1$, $\phi=0.49$, and
$\tau=1.4 \times 10^5$ at $F_p=7$ for
$N_p=50$, 100, 300, 500, 700, 900, 1100, 1300, 1500, 1700, 1900, and 2000,
from top to bottom.
The Hall angle generally starts off low and
increases with increasing drive.
In this case, the pin-free value is $\theta_H^{\rm int} = 53^\circ$.
}
\label{fig:5}
\end{figure}

\begin{figure}
\includegraphics[width=\columnwidth]{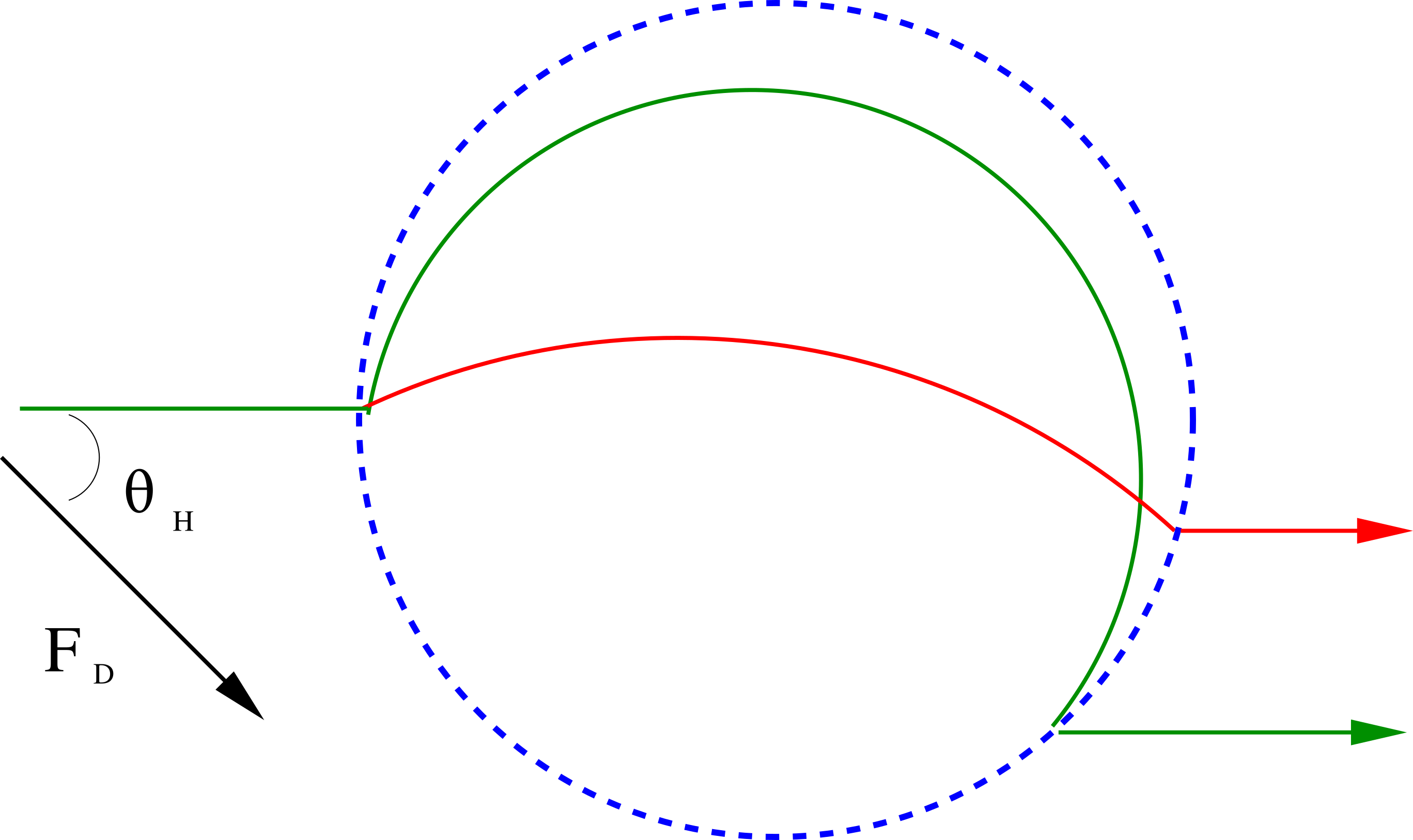}
\caption{
Schematic showing the motion of a chiral particle through a pinning
site (dashed circle), where the frame of reference has been rotated so
that the pin-free Hall angle points directly to
the right.
The driving force $F_D$ (black arrow) is at a downward angle with respect to
$\theta_H^{\rm int}$.
The green line shows the trajectory
for a small driving force, where the Magnus effect on the pinning
interaction creates a large side jump toward the direction of the drive,
reducing the average Hall angle.
The red line is the trajectory of a fast moving particle that
spends little time in the pinning site and
only experiences a small side jump.
}
\label{fig:New}
\end{figure}

We next introduce quenched disorder in the form of randomly placed
pinning sites and apply a finite drift force $F_D$.
In non-active systems such as skyrmions under the same conditions,
the Magnus term causes the particles 
to move at a finite Hall angle with respect to the drift direction.
In Fig.~\ref{fig:4} we plot
$\langle V_{x} \rangle$ and
$\langle V_{y} \rangle$ versus $F_{D}$ in a sample with
$\omega=1.1$, $\phi=0.49$, and $\tau=1.4\times 10^5$
at $F_p=7$
for varied numbers of pinning sites $N_{p}$ ranging from $N_p=50$ to $N_p=2000$.
For this value of $\omega$, a single particle in the
absence of pinning moves at the intrinsic Hall angle of
$\theta_H^{\rm int} = \arctan(\omega/\alpha_d) = 53^\circ$
with respect to the driving direction.
We find that for $N_{p}= 50$,
the velocity curves increase linearly with $F_D$, while
$\langle V_x \rangle$ and $\langle V_y \rangle$ are close together
with $\langle V_y\rangle$ sitting slightly
above $\langle V_x\rangle$.
As $N_{p}$ increases, both velocities drop but
$\langle V_x \rangle$ decreases more slowly
than $\langle V_y \rangle$.
At large values of $N_{p}$, the velocity-force 
curves become increasingly nonlinear,
similar to what is observed in
other driven systems with strong pinning \cite{Reichhardt17}.
In Fig.~\ref{fig:5}, we plot the measured Hall angle
$\theta_H  = \arctan(\langle V_y \rangle/\langle V_x \rangle)$
versus $F_{D}$ for the system from
Fig.~\ref{fig:4}.
When $N_{p} \leq 100$, $\theta_{H}$ is close to
the intrinsic value $\theta_H^{\rm int}$,
but for larger values of $N_{p}$,
$\theta_H$ has a low value for low drives and
gradually increases with increasing $F_{D}$.
For example, at $N_{p}= 700$, $\theta_{H}\approx 23^\circ$
at the lowest drives and increases up to
$\theta_H \approx 40^\circ$ at the highest drives.

\begin{figure}
\includegraphics[width=\columnwidth]{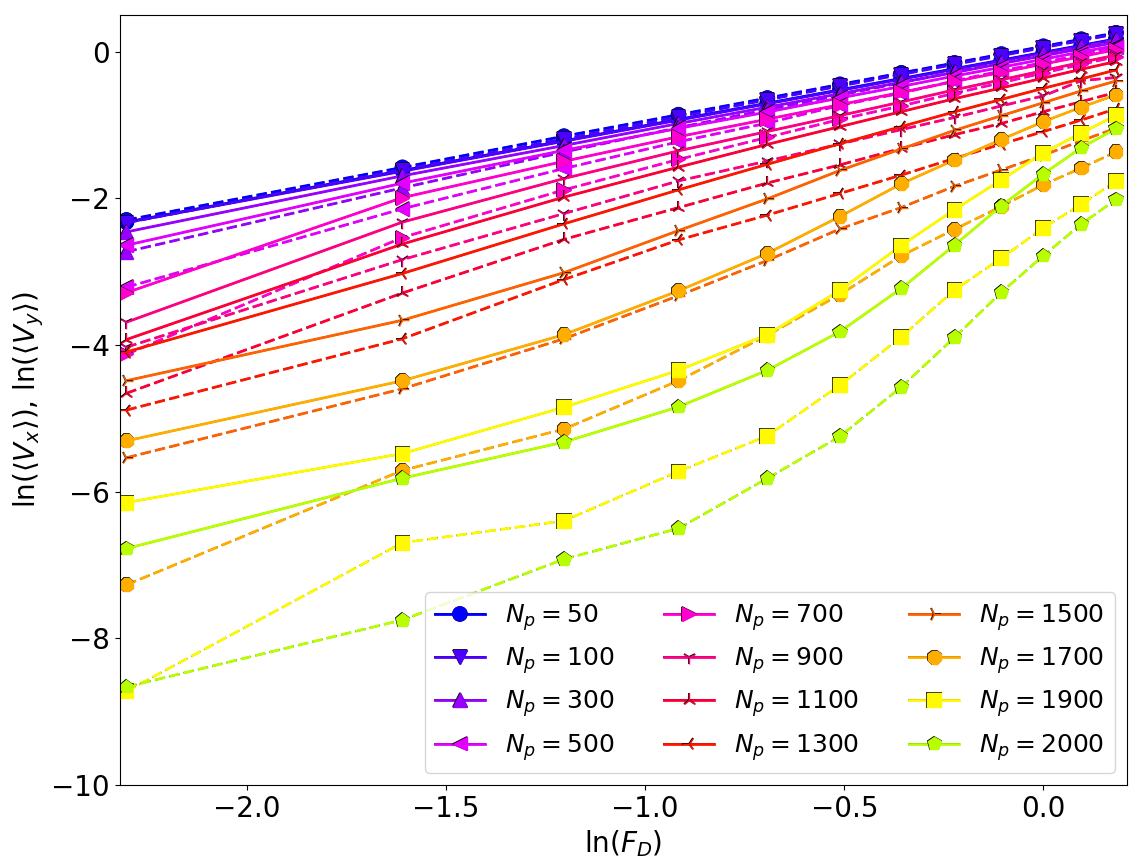}
\caption{$\langle V_x \rangle$ (solid lines) and
$\langle V_y \rangle$ (dashed lines) vs $F_{D}$ curves
on a log-log scale for the system in Fig.~\ref{fig:4} with
$\omega=1.1$, $\phi=0.49$, and $\tau=1.4 \times 10^5$ at
$F_p=7$ for $N_p=50$, 100, 300, 500, 700, 900, 1100, 1300, 1500, 1700,
1900, and 2000, from top to bottom.
There is a linear increase in the velocity with increasing drive
for small values of $N_{p}$.
For larger values of $N_p$, the increase becomes
non-linear and there is a region where
$\langle V_{x,y} \rangle \propto F^{\alpha}_{D}$
with exponents of $1.5 < \alpha < 2.0$.
 }
\label{fig:6}
\end{figure}

The drive dependence of the Hall angle
that we find for the chiral disk system
is very similar to the drive dependence
observed for skyrmions driven over random pinning arrays.
In the skyrmion case, the Hall angle is reduced at low drives
because the skyrmion undergoes a side jump every time it passes
through a pinning site as a result of the Magnus term
\cite{Reichhardt15, Jiang17, Litzius17, Fernandes20, Reichhardt22a}.
In Fig.~\ref{fig:6}, we
show a schematic of the side jump effect where the frame of reference
has been rotated so that the intrinsic Hall angle points directly to the
right.
When the particle enters the pinning site,
it experiences a radial force from the pinning that pushes it toward the
center of the pin. The Magnus force rotates this radial force partially
into the perpendicular direction,
causing the particle to move in an arc.
For smaller drives, the particle is shifted downward by a large amount
toward the driving direction, and if the particle repeatedly strikes
pinning sites as it moves,
its average Hall angle
is markedly reduced compared to the pin-free value.
At high $F_{D}$, the size of the side jump diminishes since
the amount of time the particle spends interacting with the pinning
site is reduced.
The results in Fig.~\ref{fig:5} indicate
that drifting active spinners should also experience
a Hall effect and, in the presence of pinning,
should show a drive dependent Hall angle
similar to that found for skyrmions.

The closest system to what we have studied
appears in the work of Lou {\it et al.}
\cite{Lou22}, who considered nonactive spinning
particles moving through an obstacle array.
In that study, the spinning particles
had no Hall angle in the absence of obstacles,
and a finite Hall angle appeared only when
the spinning particles collided with obstacles, which
produced a transverse shift in the particle positions.
In our case, the Magnus force also acts on the drift force,
so there is a finite Hall angle even for zero pinning,
while interaction with pinning
causes a change in the Hall angle.

In Fig.~\ref{fig:6}, we replot the
$\langle V_x \rangle$ and $\langle V_y \rangle$ versus $F_{D}$ curves
from Fig.~\ref{fig:4}
on a log-log scale.
When $N_{p} < 1500$, the velocity-force curves
are linear over most of their range with
$\langle V \rangle \propto F_{D}$. For these 
smaller values of $N_p$, most of the particles continue to move
between pinning sites since $N \gg N_p$, and the particles that do
become trapped at pins serve as localized drag sites that increase
the effective viscosity of the particles flowing around them.
For $N_{p} \geq 1500$, three regimes emerge.
At low $F_{D}$ and high $F_D$, the velocities increase linearly with the drive,
while at intermediate $F_D$ there is a regime in which
$\langle V_{x,y} \rangle \propto (F_{D})^\alpha$ with $1.5 < \alpha < 2.0$.
In previous work on two-dimensional driven systems with strong disorder,
a power law dependence of the velocity on the drive was observed with
different exponents of $\alpha<1.0$ for the elastic flow regime
and 
$1.3 < \alpha < 2.0$ in the plastic flow regime
\cite{Reichhardt17}.
In our system, at low drives the particles that are not trapped directly
by the pinning sites undergo interstitial flow past the pinned
particles, giving behavior similar to that of fluid flow through
a porous medium.
At intermediate drives, the pinned particles become only intermittently
pinned, leading to plastic flow, while at high drives,
all of the particles are continuously moving
and the system acts like a fluid again.
Figure~\ref{fig:5} shows 
that for $N_{p} > 1500$, the Hall angle is low and increases
linearly with drive, while for $N_{p} < 1500$,
the increase in the Hall angle is both more rapid and nonlinear.
We expect that for $F_{D}/F_{p} > 1.0$,
$\theta_H$ will approach the intrinsic value
$\theta_H^{\rm int}$ since
the particles would be moving so rapidly that they could only
interact weakly with the pinning sites.

\begin{figure}
\includegraphics[width=\columnwidth]{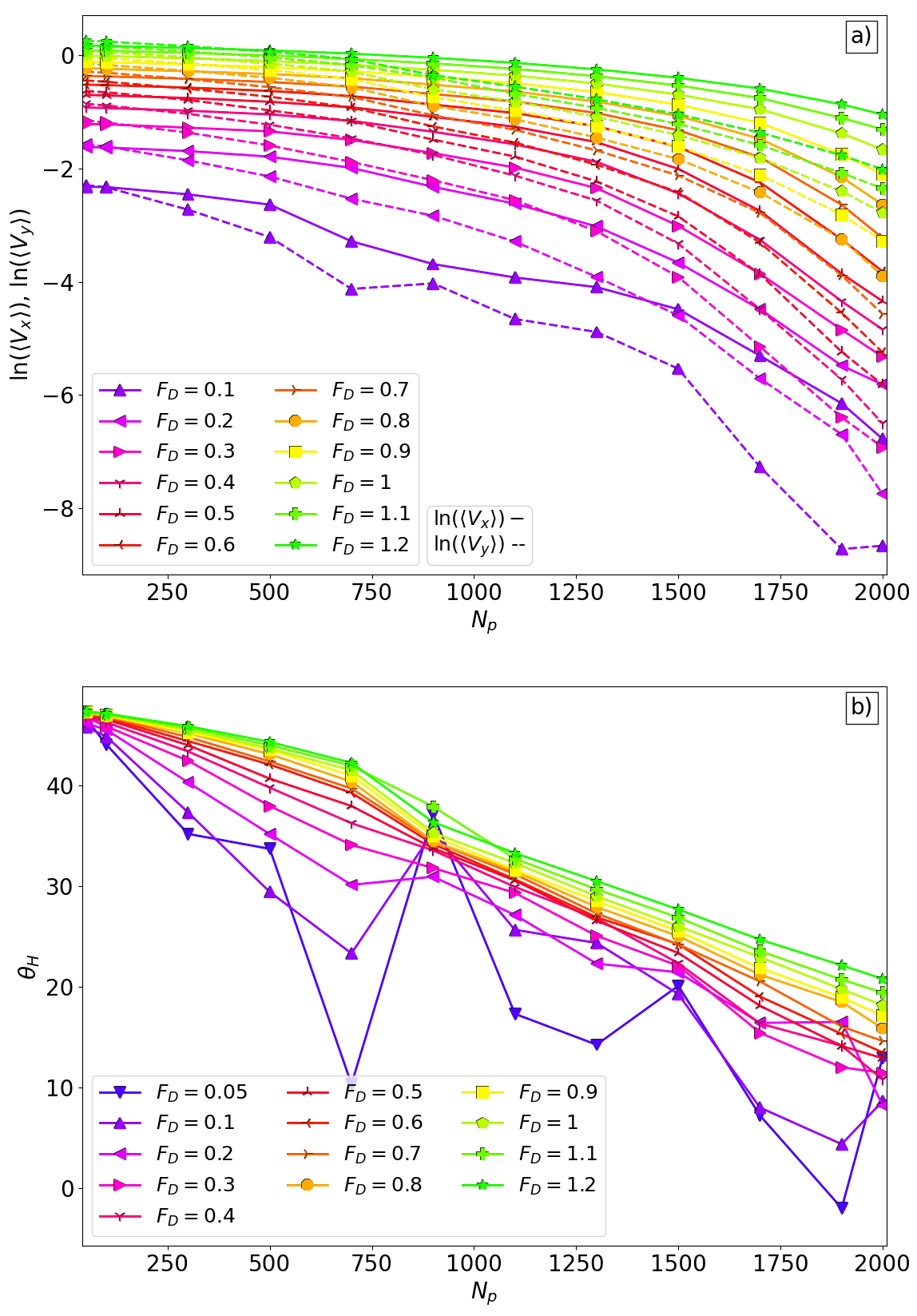}
\caption{
(a) $\langle V_x \rangle$ (solid lines) and $\langle V_y \rangle$ (dashed
lines) vs $N_{p}$ on a semilog scale for the system from
Fig.~\ref{fig:4} with $\omega=1.1$, $\phi=0.49$, and
$\tau=1.4 \times 10^5$ at $F_p=7$ for
$F_D=0.1$, 0.2, 0.3, 0.4, 0.5, 0.6, 0.7, 0.8, 0.9, 1.0, 1.1, and 1.2
from bottom to top.
(b) The corresponding $\theta_H$ vs $N_{p}$ curves on a linear scale
showing that there is a reduction of the Hall angle
with increasing $N_{p}$ and decreasing $F_{D}$.
}
\label{fig:7}
\end{figure}

In Fig.~\ref{fig:7}(a)
we plot $\langle V_x \rangle$ and $\langle V_y \rangle$
versus $N_{p}$ on a semilog scale for the system
from Fig.~\ref{fig:6} at $\rho = 0.49$, $\omega = 1.1$,
$\tau=1.4\times 10^5$, and $F_{p} = 7.0$
for varied $F_{D}$.
The velocities decrease with increasing $N_p$, and the decrease in
$\langle V_y \rangle$ is more pronounced than that of
$\langle V_x \rangle$.
The corresponding plot of
$\theta_H$ versus $N_p$ in Fig.~\ref{fig:7}(b) shows that
$\theta_H$ decreases as the number of pinning sites
increases, indicating that there is increased
scattering by the pinning sites as $N_p$ becomes larger.
Although the $\theta_H$ measurements become noisy at low
values of $F_D$, the clear general trend
is that the Hall angle decreases with decreasing $F_{D}$ and
increasing $N_{p}$.

\begin{figure}
\includegraphics[width=\columnwidth]{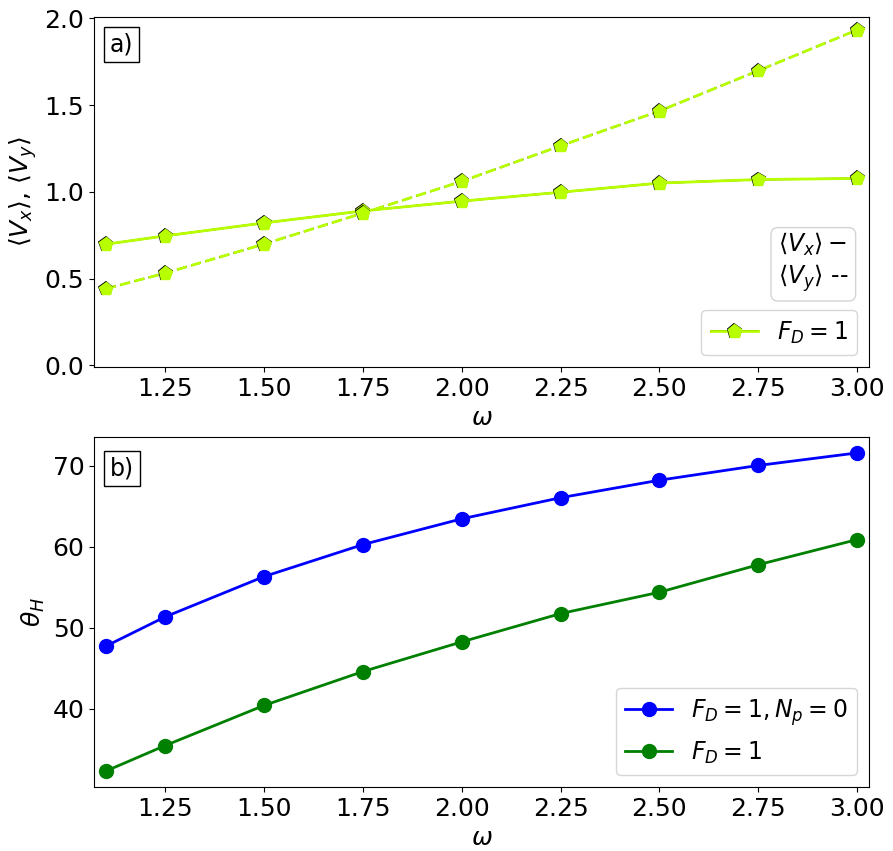}
\caption{
(a) $\langle V_x \rangle$ (solid line) and $\langle V_y \rangle$ (dashed line)
vs Magnus strength $\omega$ 
at a fixed drive of $F_{D} = 1.0$ for
a system with $F_{p}= 7.0$, $N_{p}= 1100$,
$\rho = 0.49$ and $\tau = 1.4\times 10^5$.
(b) The corresponding $\theta_H$ (green)
along with
the expected value of $\theta_H^{\rm int}$ in a pin-free system (blue)
vs $\omega$, both for $F_D=1$.
}
\label{fig:8}
\end{figure}

In Fig.~\ref{fig:8}(a) we plot
$\langle V_x \rangle$ and $\langle V_y \rangle$ at
a fixed drive of $F_{D} = 1.0$
for a sample with $F_{p}= 7.0$, $N_{p}= 1100$,
$\rho = 0.49$, and $\tau = 1.4\times 10^5$.
As the Magnus strength $\omega$ increases,
$\langle V_y \rangle$ increases more rapidly than
$\langle V_x \rangle$.
Figure~\ref{fig:8}(b) shows the corresponding
measured $\theta_H$ versus $\omega$ along with the
expected value $\theta_H^{\rm int}$
versus $\omega$ for a pin-free system with $N_p=0$.
We find that the Hall angle is always reduced by the pinning,
$\theta_H<\theta_H^{\rm int}$,
due to the asymmetric scattering of the particles by
the pinning sites.

\begin{figure}
\includegraphics[width=\columnwidth]{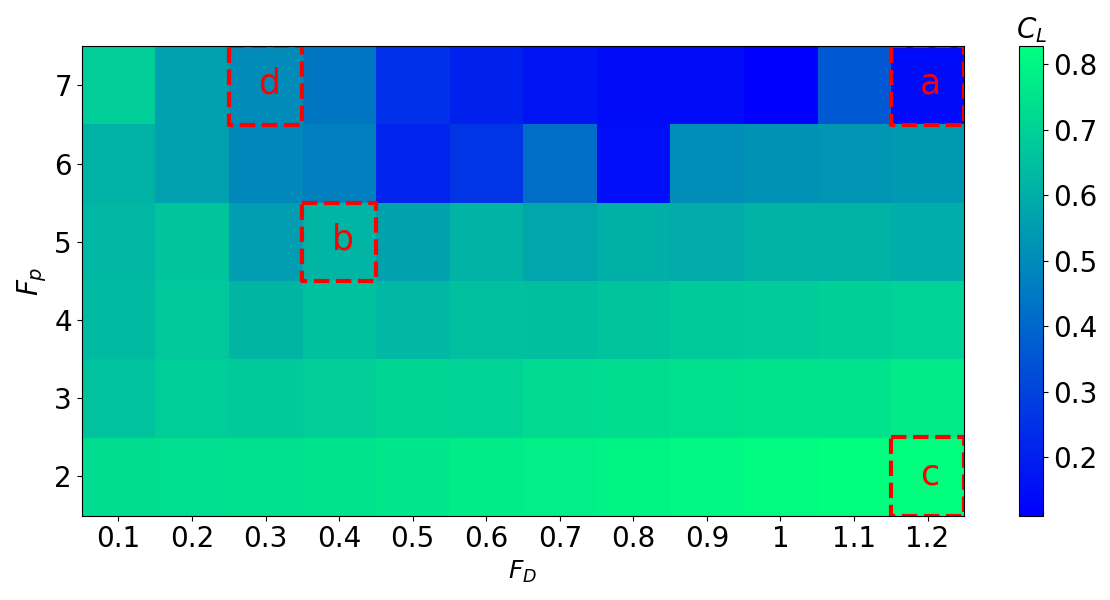}
\caption{Heat map of the fraction $C_L$ of particles in the
largest cluster as a function of $F_p$ vs $F_{D}$ for
samples with
$\phi = 0.49$, $\omega = 1.1$, $N_{p}= 900$, and
$\tau = 1.4\times 10^5$.
The red squares indicate the points at which the images in
Fig.~\ref{fig:10} were obtained.
}
\label{fig:9}
\end{figure}

\begin{figure}
\includegraphics[width=\columnwidth]{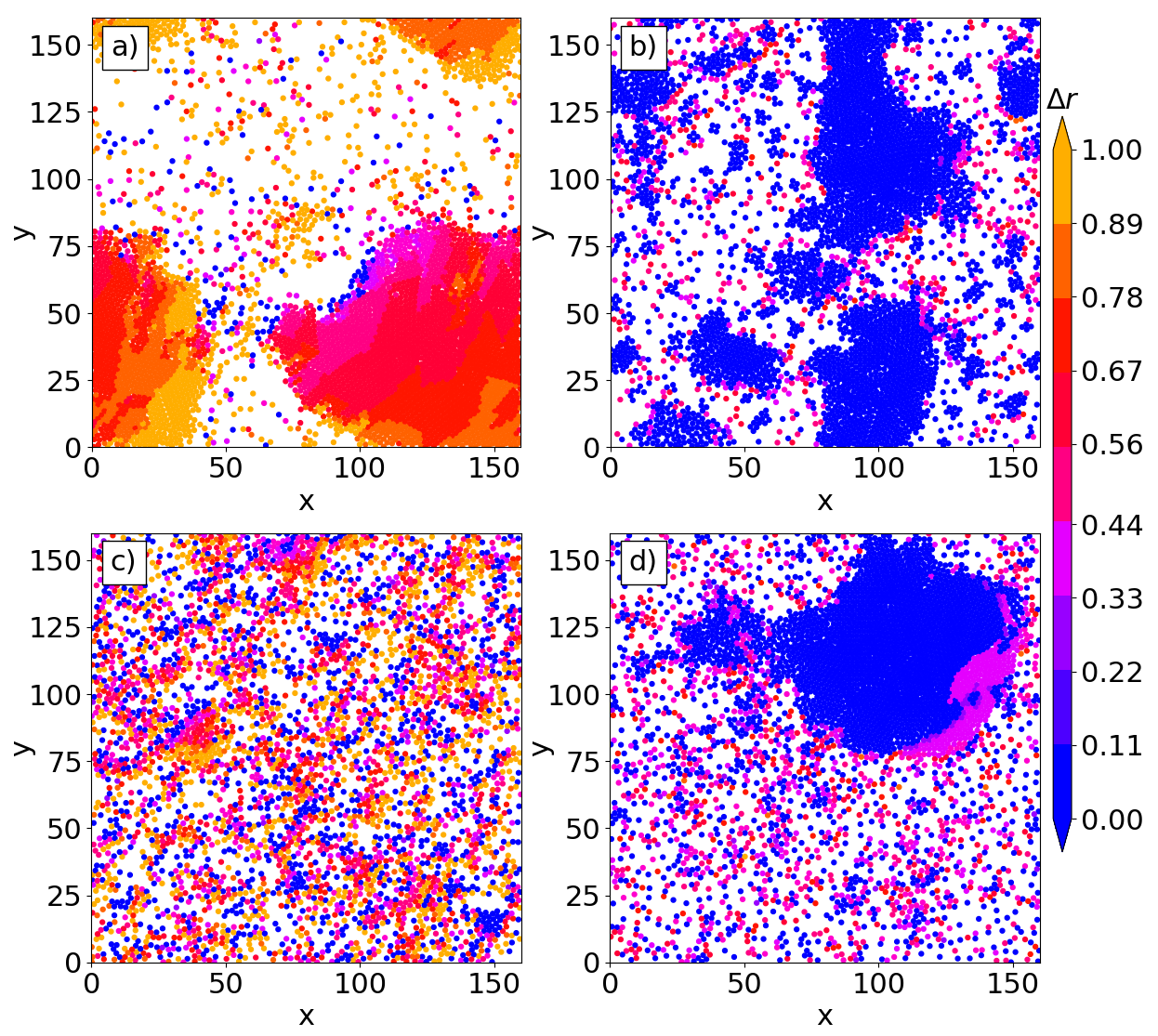}
\caption{
Images of particle configurations in
the system from Fig.~\ref{fig:9} at $\phi=0.49$, $\omega=1.1$, $N_p=900$, and
$\tau=1.4 \times 10^5$. Colors indicate the net displacement
$\Delta r$ of the particle during a period of 500 simulation
time steps. 
(a) $F_{p}= 2$ and $F_{D} = 1.2$, where there is strong clustering.
(b) $F_{p} = 7$ and $F_{D} = 0.1$.
(c) $F_{p}= 7$ and $F_{D} = 1.2$. 
(d) $F_{p}= 5$ and $F_{D} = 0.2$.
}
\label{fig:10}
\end{figure}

In Fig.~\ref{fig:9} we plot a heat map of the fraction $C_L$ of
particles in the largest cluster as a function of
$F_{p}$ versus $F_{D}$ for a system with
$\phi = 0.49$, $\omega = 1.1$, $N_{p}= 900$, and
$\tau = 1.4\times 10^5$.
For low pinning forces,
the system is phase separated and $C_L$ is large for all
drives. An image of the clustered state obtained for $F_p=2$ and
$F_D=1.2$ appears in Fig.~\ref{fig:10}(a).
The particles in the low density region that have very small
velocities have become trapped at pinning sites.
Since the system is subjected to a finite drift force, it is
more difficult to detect the presence of an edge current using the
net displacement measure $\Delta r$; however, an edge current is still
present, and the entire cluster
acts like an effective macroscopic
particle and moves at a Hall angle with respect
to the external drive.
When $F_{p}$ is raised to a large value,
Fig.~\ref{fig:9} indicates that a cluster state appears only
for $0 < F_{D} < 0.2$,
as illustrated in Fig.~\ref{fig:10}(b) for a sample with
$F_{p}= 7$ and $F_{D} = 0.1$.
Here the system forms multiple smaller clusters that are each
surrounded by an edge current.
As $F_D$ increases at high $F_p$, $C_L$ drops when the clusters begin to break
apart into the plastically flowing state, shown in Fig.~\ref{fig:10}(c)
for a sample with $F_p=7$ and $F_D=1.2$.
If $F_p$ is lowered to an intermediate value, the clustering persists
over a wide drive range but is somewhat suppressed at intermediate
drives. Figure~\ref{fig:10}(d) shows that the system has reformed
a cluster for $F_p=5$ and $F_D=0.2$; however, the number of pinned particles
is substantially higher than in the $F_p=2$ and $F_D=1.2$ sample
from Fig.~\ref{fig:10}(a).

The heat map of $C_L$ in Fig.~\ref{fig:9}
indicates that for low $F_{D}$, the system forms a cluster state at
high $F_{p}$, while for higher $F_{D}$,
at high $F_{p}$ the clusters are broken apart by plastic flow.
We expect that the system would form a strongly clustered state again
even for $F_{p} = 7.0$ once $F_{D} > F_{p}$.
We can compare our results to previous simulations
performed with $\omega = 0$, where it was also observed
that the MIPS cluster state breaks up for strong pinning
at intermediate drive but can reappear at higher drives \cite{Sandor17a}. 
This is also consistent with studies in
nonactive particles driven over strong quenched disorder,
which showed that the system is the most strongly disordered
for intermediate drives but can dynamically order at higher drives
\cite{Reichhardt17}.
Whether or not the system is able to dynamically reorder
should also depend on the type
of quenched disorder present.
For the pinning sites that we consider here,
a strong drive causes the particles to move rapidly over the pinning
and decouple from it;
however, if the pinning sites are replaced by obstacles,
the particles remain in a disordered configuration
even for very high drives. 

\section{Non-Active Chiral Disks}

\begin{figure}
\includegraphics[width=\columnwidth]{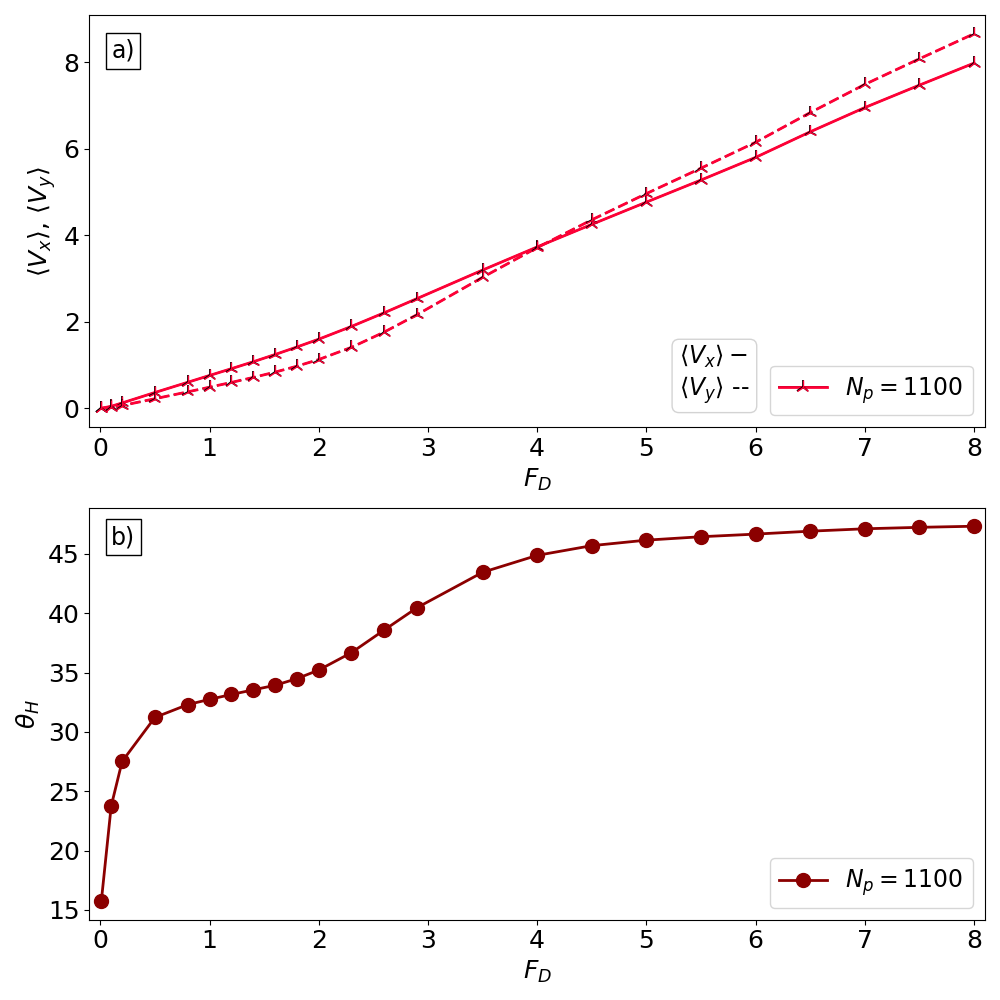}
\caption{
(a) $\langle V_x \rangle$ (solid line) and
$\langle V_y \rangle$ (dashed line) vs $F_{D}$ for particles
with a Magnus term but no activity, $F_m=0$,   
in a system with $\phi = 0.49$,
$\omega = 1.1$, and $N_{p} = 1100$.
(b) The corresponding Hall angle
$\theta_H$ vs $F_D$ asymptotically tends
toward the pin free value of $\theta_H^{\rm int}=53^\circ$.
}
\label{fig:11}
\end{figure}

\begin{figure}
\includegraphics[width=\columnwidth]{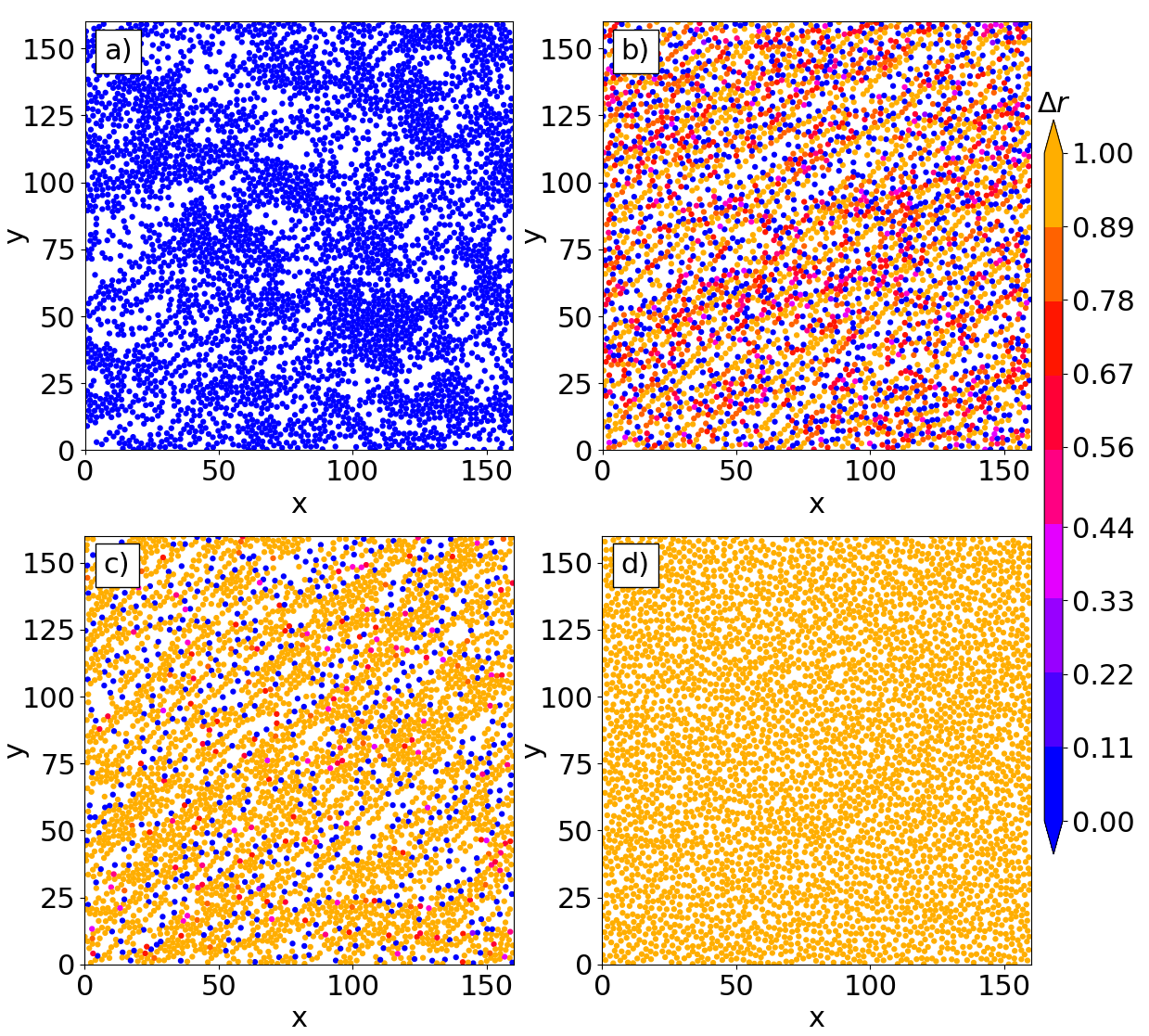}
\caption{
Images of particle configurations in the system from Fig.~\ref{fig:11}
with no activity, $F_m=0$, at $\phi=0.49$, $\omega=1.1$, and $N_p=1100$.
Colors indicate the net displacement $\Delta r$ of the particle during a
period of 500 simulation time steps.
(a) $F_{D} = 0.1$. (b) $F_{D}= 1.5$. (c) $F_{D} = 3.0$.
(d) $F_{D} = 7.0$.}
\label{fig:12}
\end{figure}

To examine how general the drive dependence of the
Hall effect is, we consider a system in which the particles
have a Magnus term but no motor force, $F_m=0$.
In Fig.~\ref{fig:11}(a) we plot $\langle V_x \rangle$
and $\langle V_y \rangle$ versus $F_{D}$
for an $F_m=0$ system with $\phi = 0.49$, $\omega = 1.1$,
$F_{p}= 7.0$, and $N_{p} = 1100$, the same as one of the samples
from Figs.~\ref{fig:4} to \ref{fig:8} but with no activity.
We find that
$\langle V_x \rangle$ increases more rapidly
than $\langle V_y\rangle$ with increasing
$F_D$ at lower drives.
The corresponding plot of
$\theta_H$ versus $F_{D}$ in Fig.~\ref{fig:11}(b)
shows that $\theta_H$ has a small value at low drives
and increases with increasing $F_D$ in a manner similar to that found
for the active system in Fig.~\ref{fig:5}.
In Fig.~\ref{fig:12}(a) we illustrate the particle positions and
displacements for the $F_m=0$ system from Fig.~\ref{fig:11} at
a drive of $F_{D}= 0.1$.
Most of the particles are pinned and we observe the formation of a
locally clogged state.
At $F_D=1.5$ in Fig.~\ref{fig:12}(b),
the particles travel at an angle with respect to the direction
of the external drive and a wide range of velocity values are present,
with some of the particles remaining pinned and other particles moving
rapidly through the interstitial regions.
Figure~\ref{fig:12}(c) shows the same system
at $F_{D} = 3.0$ where the velocity distribution has become strongly
bimodal, with a smaller set of pinned particles and a large fraction of
moving particles all traveling with the same speed. The Hall angle of
the motion is larger compared to Fig.~\ref{fig:12}(b).
At $F_D=7.0$ in Fig.~\ref{fig:12}(d), $F_D=F_p$ so all of the particles
are moving and a disordered uniform state emerges.
This result shows that the drive dependence of the Hall angle appears
both with and without the active motor force as long as the equations of
motion contain a Magnus term.

\section{Summary} 

We have investigated
a two-dimensional system of active run-and-tumble particles that have
both a damping and a Magnus term in their dynamics.
The damping aligns the particle velocities with the net force exerted
on the particle, while the Magnus term contributes a velocity component
that is perpendicular to the net force.
For densities and running times at which
the system forms a motility-induced phase separated state (MIPS)
in the absence of the Magnus term,
we find that inclusion of a Magnus term produces
a chiral MIPS phase containing a large rotating cluster
with an edge current. The direction of rotation and of the edge
current
matches the chirality of the Magnus term.
As the size of the Magus term increases, we
observe a shear banding effect
in which particles along the edges of the cluster
form multiple bands moving at different speeds
while the center of the cluster rotates as a rigid crystalline solid.
We also find that the Magnus term stabilizes the MIPS state down to
lower run lengths compared to a Magnus-free system because
the Magnus forces push particles on the cluster boundary that would otherwise
move radially outward away from the cluster into trajectories that spiral
around the edge of the cluster, keeping the particles
localized close to the cluster.
Similar Magnus-induced clustering effects have been observed in
systems of magnetic skyrmions.
As the Magnus term increases, the shear bands increase in width
and destroy the crystalline order of the cluster center,
and when the Magnus force becomes dominant,
the phase separation is lost
and a chiral labyrinthine state emerges.

We examined the effect of including quenched disorder and a uniform
drift force.
Without quenched disorder, the rotating MIPS cluster moves at a Hall
angle with respect to the applied drift force that is
determined by the ratio of the Magnus term to the damping term.
When quenched disorder is present,
nonlinearity appears in the velocity-force curves
and the Hall angle becomes drive dependent, taking a small value
at low drives and asymptotically approaching the intrinsic or pin-free
value as the drive increases.
This is the result of the side jump, experienced by the particles as they
move through the pinning sites, that becomes smaller as the drive becomes
larger.
The velocity dependent Hall angle
is the same effect observed for
magnetic skyrmions, which have a Magnus term in their dynamics,
driven over quenched disorder.
We find that even when the quenched disorder is very strong,
the particles still form a cluster or phase-separated state at zero drive,
but at finite drives, plastic events produced by the interaction between
pinned and moving interstitial particles tear the clusters apart;
however, at large drives, the clusters can reform.
The velocity dependence of the Hall angle persists even when the motor force
is shut off so that there is no run-and-tumble component to the particle
motion.

\begin{acknowledgments}
We gratefully acknowledge the support of the U.S. Department of
Energy through the LANL/LDRD program for this work.
This work was supported by the US Department of Energy through
the Los Alamos National Laboratory.  Los Alamos National Laboratory is
operated by Triad National Security, LLC, for the National Nuclear Security
Administration of the U. S. Department of Energy (Contract No. 892333218NCA000001).
AL was supported by a grant of the Romanian Ministry of Education
and Research, CNCS - UEFISCDI, project number
PN-III-P4-ID-PCE-2020-1301, within PNCDI III.
\end{acknowledgments}

\bibliography{mybib.bib}

\end{document}